\documentclass[
10pt]{article}

\usepackage[bookmarks, colorlinks=TRUE, linkcolor=blue, urlcolor=blue,
citecolor=black, pdftitle={Sparse Network  Estimation for  Dynamical Spatio-temporal Array Models}, pdfauthor={Adam Lund}]{hyperref}

\RequirePackage[OT1]{fontenc}
\RequirePackage{ mathtools}
\newcommand{\tr}{\mathrm{tr}}

\usepackage[textsize=tiny, backgroundcolor=white, linecolor=red, bordercolor=red]{todonotes}
\usepackage{lineno}
\usepackage{setspace}
\usepackage{mathpazo}
\usepackage[sort&compress, authoryear]{natbib}
\usepackage{amsmath}
\usepackage{amssymb}
\usepackage{amsthm}
\usepackage{algorithm}
\usepackage{algorithmic}
\usepackage{float}
\usepackage{fixme}
\usepackage{hyperref}
\newcommand{\E}{\mathrm{E}}
 
\fxsetup{
 status=draft,
 layout=inline,
 theme=color
}

\usepackage{fixme}
\fxsetup{
 theme=color
}

\usepackage{lscape}
\usepackage{caption}
\usepackage{subcaption}
\usepackage{graphics}
\usepackage{lipsum}
\usepackage{amssymb}
\usepackage{graphicx}

\usepackage{float}
\usepackage{algorithm}
\usepackage{algorithmic}

\usepackage{mathrsfs}

\newtheorem{thm_def}{Definition}

\newtheorem{thm_prop}{Proposition}

\newcommand{\NN}{\mathbb{N}}
\newcommand{\GG}{{\mathbb G}} 
\newcommand{\RR}{\mathbb{R}}
\newcommand{\ve}{\mathrm{vec}}

\newcommand{\F}{{\mathcal F}} 

\newcommand{\Sop}{\mathrm{S}}
\newcommand{\Fop}{\mathrm{F}}
\newcommand{\Hop}{\mathrm{H}}

\newcommand{\dif}{\mathrm{d}}

\newcommand{\C}{{\mathcal C}}
\newcommand{\Hcal}{{\mathcal H}}
\newcommand{\arr}[1]{\boldsymbol{\mathrm{#1}}}
\newcommand{\ar}[1]{\boldsymbol{#1}}

\usepackage{pdfpages}

\begin{document}

\title{Sparse Network  Estimation for  Dynamical Spatio-temporal Array Models}
\author{Adam Lund        and Niels Richard Hansen 
\thanks{University of Copenhagen,
Department of Mathematical Sciences, 
Universitetsparken 5,
2100 Copenhagen \O, Denmark, e-mail: adam.lund@math.ku.dk (A. Lund).}}
\date{}
\maketitle


\begin{abstract}
Neural field models represent neuronal communication on a population level via synaptic weight functions. Using voltage sensitive dye (VSD) imaging it is possible to obtain measurements of neural fields with a relatively high spatial and temporal resolution. 
The synaptic weight functions represent functional connectivity in the brain and give rise to a spatio-temporal dependence structure. We present a stochastic functional differential equation for modeling neural fields, which leads to a vector autoregressive model of the data via basis expansions of the synaptic weight functions and time and space discretization.
Fitting the model to data is a pratical challenge as this represents a large scale regression problem. By using a 1-norm penalty in combination with localized basis functions it is possible to learn a sparse network representation of the functional connectivity of the brain, but still, the explicit construction of a design matrix can be computationally prohibitive.
We demonstrate that by using tensor product basis expansions, the computation of the penalized estimator via a proximal gradient algorithm becomes feasible. It is crucial for the computations that the data is organized in an array as is the case for the three dimensional VSD imaging data. This allows for the use of array arithmetic that is both memory and time efficient.The proposed method is implemented   and showcased in the \textsf{R} package \verb+dynamo+ available from CRAN.
 
\end{abstract}
 
%
 
\section{Introduction}\label{sec:intro}
 
Neural field models are models of  aggregated membrane voltage  of a large and spatially distributed population of neurons. The neuronal network is determined by spatio-temporal synaptic weight functions in the neural field model, and we will refer to these weight functions as the \emph{propagation network}. This network determines how signals are propagated hence it is of great interest to learn the propagation network from experimental data, which is the inverse problem for neural field models.
 
The literature on neural fields is vast and we will not attempt a review, but see \cite{Bressloff:2012b, Coombes:2014} and the references therein. The typical neural field model considered is a deterministic integrodifferential equation. The inverse problem for the deterministic Amari equation was treated in \cite{Graben:2009} and \cite{Potthast:2009}, and stochastic neural field models was, for instance, treated in Chapter 9 in \cite{Coombes:2014} and in \cite{Faugeras:2015}. One main contribution of the latter paper, \cite{Faugeras:2015}, was to treat a stochastic version of the Amari equation in the well developed theoretical framework of functional stochastic evolution equations.
 
Despite the substantial literature on neural fields, relatively few papers have dealt directly with the estimation of neural field components from experimental data. Pinotsis et al. \cite{Pinotsis:2012} demonstrated how a neural field model can be used as the generative model within the dynamical causal modeling framework, where model parameters can be estimated from electrophysiological data. The modeling of voltage sensitive dye (VSD) imaging data in terms of neural fields was discussed in \cite{chemla2010}, and Markounikau et al. \cite{Markounikau:2010} estimated parameters in a neural field model directly from VSD   data.
 
In this paper, VSD imaging data is considered as well. This \emph{in vivo} imaging technique has a sufficiently high resolution in time and space  to detect propagation of changes in membrane potential on a mesoscopic scale, see \cite{roland2006}.  A prevalent notion in the neuroscience literature is that the  network    connecting the neurons, and  through which brain signals are propagated,    is sparse, and     that the propagation exhibits a time delay.  If a spiking neuron, for instance, only affects neurons via synaptic connections to a very localized region the network is spatially sparse, while connections to remote regions result in temporal sparsity and long range dependence, see, e.g., \cite{brunel2000, sporns2004, roxin2011, bressloff2012, touboul2014}. Finally the possibility of feedback waves  in the brain, e.g., as suggested in \cite{roland2006}, could also be explained  by  spatio-temporal dynamics  depending on more than just the instantaneous past. These considerations lead us to suggest a class of stochastic neural field models that allows for time delay, and a proposed estimation methodology that provides sparse nonparametric estimation of synaptic weight functions. Thus we do not make assumptions about spatial homogeneity or isotropy of the functional connectivity, nor do we assume that the signal propagation is instantaneous.
 
In order  to derive a statistical model  that, from a computational viewpoint, is feasible for  realistically sized  data sets, a time and space discretized version of  the infinite dimensional dynamical model is  obtained by replacing the various integrals with Riemann-It\^o type summations and relying on an Euler  scheme approximation.  This approximation scheme makes it possible to derive a  statistical model with an associated likelihood function such that regularized maximum-likelihood estimation becomes computationally tractable. Especially,  we show that by expanding each component function in a tensor product basis we can formulate the statistical model as  a type of multi-component linear array model, see \cite{lund2017}.
 
The paper is organized as follows. First we give a brief technical introduction to the stochastic dynamical model that form the basis for the paper. Then we present the aggregated results from the application of our proposed estimation methodology to part of a VSD imaging data set. The remaining part of the paper presents the derivation of the linear array model and the key computational techniques required for the actual estimation of the model using array data. The appendix contains further technical proofs, a meta algorithm  and implementation details. Finally to further illustrate our results we also provide a  Shiny app  available at  \href{http://shiny.science.ku.dk/AL/NetworkApp/}{\texttt{shiny.science.ku.dk/AL/NetworkApp/}} as well as  supplementary material, \cite{lund2018b},   containing results from fitting the model to individual trials and the aggregated  result for the entire data set.
 
\section{A stochastic  functional  differential   equation}\label{sec:dynmodel}
The data that we will ultimately consider is structured as follows. With $\tau,T>0$, $\mathcal{T}\coloneqq [-\tau,T]$ and  $N_x,N_y, M,L\in \NN$ we record, to each of   $N_t \coloneqq  M + L + 1$   time points 
\begin{alignat}{4}\label{onenew}
-\tau=t_{-L}<\ldots<t_0< \ldots<t_{M}=T,
\end{alignat}
  a 2-dimensional rectangular $N_x\times N_y$ image of  neuronal activity  in an area of the brain represented by the Cartesian product $\mathcal{S}\coloneqq \mathcal{X}\times\mathcal{Y}\subseteq \RR^2$. These images consist of  $D\coloneqq N_xN_y$  pixels each represented by a coordinate $(x_i,y_j)$ lying on a grid $\GG^2\subseteq \mathcal{S}$. To each time point each pixel has a color  represented by  a value $\ar v(x_{i},y_{j},t_k)\in \RR$. Thus the observations are  naturally  organized  in a 3-dimensional array $\ar{v}\coloneqq \{\ar v(x_{i},y_{j},t_k)\}_{i,j,k}$ where the first  two  dimensions   correspond to the  spatial dimensions and the third dimension corresponds to the  temporal dimension.
 
   As such it is natural to  view $\ar v$ as a discretely observed  sample in time and space of an underlying  spatio-temporal  random field  $\ar V$.  Following  Definition 1.1.1 in \cite{adler2009} any measurable map $\ar{V}:\Omega\to \RR^{\mathcal{R}}$, with $\mathcal{R}\subseteq\RR^d, d\in \NN$, a parameter set,    is called a $(d,1)$-random field or simply a $d$-dimensional random field.  Especially,   for the brain image data, $\ar V$ is real valued with  a 3-dimensional parameter set $\mathcal{R}\coloneqq \mathcal{S}\times\mathcal{T}$ where   $\mathcal{S}$ refers to space while $\mathcal{T}$ refers to time.  We emphasize the conceptual asymmetry between these dimensions by   calling $\ar V$ a spatio-temporal random field. For fixed $t$, as  $\ar V(t)\coloneqq \ar V(\cdot,\cdot,t):\RR^2\to\RR$  this model will inevitably be a  stochastic dynamical model on a function space, that is an infinite dimensional stochastic dynamical model. 
 
Following the discussion in the introduction above we propose to model the random neural field $\ar V$ via a stochastic functional differential equation (SFDE) with the propagation network incorporated into the drift as a spatio-temporal linear filter (a convolution) with an impulse-response  function (convolution kernel) quantifying the network. The solution to the SFDE in a Hilbert space $\Hcal$ is then the underlying time and space continuous model $\ar V$ for the data $\ar v$.
 
To introduce the  general model more formally let $(\Omega, \F, \Pr)$ be a probability space endowed with an increasing and right continuous family $(\F_t)$ of complete sub-$\sigma$-algebras of $\F$.  Let $\Hcal $ be a reproducing kernel Hilbert space (RKHS) of continuous functions over the compact set $\mathcal{S}\times \mathcal{S}$.
Suppose  $(\ar V(t))_t$ is a continuous $\F_t$-adapted, $\Hcal $-valued stochastic process and
  let $\C\coloneqq \C([-\tau,0],\Hcal )$ denote the Banach space of continuous maps from  $[-\tau,0]$ to $\Hcal $.  Then   $(\ar V_{t})_{t}$, where
\begin{alignat}{4}
\ar V_{t} \coloneqq\{\ar V(t+s)\}_{s\in (-\tau,0)}, \quad t\geq0,  
\end{alignat}
defines    a $\C$-valued stochastic process  over  $\RR_+$. We call    $\ar V_{t}$  the  $\tau$-{memory} of the random field  $\{\ar V(r)\}_{r\in \mathcal{S}\times \mathcal{T}}$ at  time $t\geq0$.
 
  Let $\mu:\C\times[0,T]\to \Hcal $ be a bounded linear operator and  consider  the stochastic functional differential equation (SFDE) on $\Hcal $ given by
\begin{alignat}{4}\label{eq1}
\dif \ar V(t)=\mu(\ar V_{t},t)\dif t+\dif \ar W(t).
\end{alignat}
Here $\ar W$  is a spatially homogenous Wiener process with spectral measure $\sigma$ ($\sigma$ is a finite symmetric measure on $\RR^2$) as in \cite{peszat1997}. That is, $\ar W$ is  a centered Gaussian random field such that $ \{W ( x,y, t)\}_t $ is a $(\F_t )_t$-Wiener process for every $(x,y) \in \RR^2$, and  for $   t, s \geq 0$ and $  (x, y), (x', y')\in  \RR^2 $
\begin{alignat}{4}\label{eq4}
  \E \{\ar W(x,y,t)\ar W(x,'y',s)\} = (t \land s)c (x-x',y-y' ).
\end{alignat}
Here  $c:\RR^2\to\RR$, the covariance function defined as the Fourier transform of the spectral measure  $\sigma$, quantifies the   spatial correlation in $\ar W$.
 
Compared to a typical SDE, the infinitesimal dynamic at time $t$ as described by \eqref{eq1}   depends on the  past via the $\tau$-memory of $V$ in the drift operator $\mu$. Hence  processes satisfying \eqref{eq1} will typically be non-Markovian.   We  note that all the memory in the system is modelled by the drift $\mu$.
 
The non-Markovian property  makes  theoretical results regarding existence,  uniqueness  and stability of solutions to \eqref{eq1} much less accessible.  Corresponding  to Section 0.2  in \cite{daprato2014}, in order to obtain theoretical results, it should be possible  to lift the  equation \eqref{eq1} and obtain  a Markovian SDE  on the Banach space $\C$. Consequently an unbounded linear operator (the differential operator) then appears in the drift.  It is outside the scope of this paper to pursue a discussion of the theoretical properties of \eqref{eq1}.  General  theoretical results  on SDEs on Banach spaces are not abundant, see,  e.g., \cite{cox2012}, where SDEs and especially SDDEs on Banach spaces are treated. Especially, Corollary 4.17  in  \cite{cox2012}  gives an  existence result for a strong solution to \eqref{eq1}.  Also  in \cite{xu2012} a mild existence results are given for an SDDE on a  Hilbert space and for the specification introduced next this result can be strengthened to a strong solution result. In general, the requirements for a solution to exist is, corresponding to the finite dimensional case, that the coefficient operators are Lipschitz continuous, which, e.g., the integral operator  presented next satisfies.
 
\subsection{Drift  operator}
The    idea here is that  the drift operator $\mu$   will capture  both external input to the system (the brain in our context) as well as the   subsequent propagation of this input  over time and space.  By decomposing the drift operator we obtain drift components responsible for modelling  instantaneous  effects and propagation effects respectively. To this end we will  specify the drift operator  by the  decomposition
\begin{alignat}{4}
  \mu(\ar V_t,t)\coloneqq \Sop(t)+\Fop(\ar V_t) + \Hop\{\ar V(t)\} .
\end{alignat}
Here $\Sop:\mathcal T\to  \Hcal $, $\Sop(t)(x,y)\coloneqq s(x,y, t)$, with $s\in L^2(\RR^3,\RR)$ a smooth function of time and space,  models a  deterministic time dependent external  input to the system. $\Hop: \Hcal  \to  \Hcal $, $\Hop\{\ar V(t)\}(x,y)\coloneqq h(x, y)\ar V(x,y,t)  $  where $h\in \Hcal $, a smooth function of  space, captures the short range (infinitesimal)  memory in the system.
 
The long range memory responsible for propagating the input to the system over time and space  is   modelled by the operator  $\Fop:\C([0,\tau],\Hcal )\to \Hcal $  given as the  integral operator 
\begin{alignat}{4}
\Fop(\ar  V_t)(x,y)=\int_{\mathcal{S}}  \int_{-\tau}^0w(x,y,x',y',r)\ar V(x',y',t+r)\dif r\dif x'\dif y'.
\end{alignat}
Here  $w\in L^2(E^5,\RR)$  is a smooth weight function quantifying the impact of previous states on the current change in  the random field $\ar V$. Especially, the value $w(x,y,x',y',r)$ is the weight by which the change in the field at location $(x,y)$ is impacted by the level of the field at location $(x',y')$ with delay  $r$. With this specification, a  solution to  \eqref{eq1}, if it exists, can  be written in integral form as
\begin{alignat}{4}\label{eq7}
   \ar V(x, y, t)=\int_0^t \bigg\{  s(x,y,u)&+\int_{\mathcal{S}}   \int_{-\tau}^0w(x,y,x',y',r)  \ar V(x',y',u+r)\dif r\dif x'\dif y' \nonumber\\
   &+h(x,y)\ar V(u)  \bigg\}\dif{}u
     +\int_0^t \dif \ar W(x, y,u) \dif{}u.
\end{alignat}
Thus \eqref{eq7}   characterizes  a solution  to a stochastic delay differential equation (SDDE) on $\Hcal $ with   delays  distributed over time and space  (spatio-temporal distributed delays)  according to  the  impulse-response  function $w$.  We can think of $w$ as quantifying a spatio-temporal network in the brain that  governs how the propagation of the  input is to be  distributed over time and space, and we will refer to $w$ as the network function.
 
Standard neural field models usually include a non-linear transformation of $\ar V$, via a so-called gain function, inside the integral operator $\Fop$. It is possible to include a known gain function transformation in \eqref{eq7} without substantial changes to our proposed methodology for estimating $s$, $w$ and $h$. However, to the best of our knowledge the appropriate choice of gain function for empirical data has not been settled. It is, of course, possible to attempt to estimate the gain function from data as well, but to avoid complicating matters we proceed by regarding \eqref{eq7} as corresponding to a linearization of the unknown gain function.  We note that the interpretation of $w$ will always be relative to the choice of gain function, and it might not quantitatively represent physiological properties of the brain, if the gain function is misspecified.
 
Next we   present an example where  a statistical model based on   the spatio-temporal  SDDE model proposed above is fitted to   real high dimensional brain image data. The derivation of this statistical model relies on a space time discretization   and is discussed   in Section \ref{sec:plam}. We note that key elements in our approach involves expanding the network function  (along with the other  component functions) using   basis functions   with compact support in time and space domain. We then apply regularization techniques  to obtain a sparse  (i.e., space-time localized) estimate of the network.
 \section{Brain imaging data}\label{sec:vsd}
The  data  considered in this section consists of  in vivo recordings of the visual cortex  in ferret brains provided by Professor Per Ebbe Roland.  In the experiment producing the data,  a voltage sensitive dye (VSD) technique was used to record a sequence of images of  the visual cortex, while presenting a   visual stimulus, a stationary white square displayed for 250 ms on a grey screen, to the ferret.   Each recording or trial  is thus  a film  showing activity in a live ferret brain before, during and after a visual stimulus is presented. The purpose of the experiment was to study   the response in  brain activity   to the stimulus and its propagation over time and space.
 
For each of a total of 13 ferrets the experiment was  repeated several times  producing a   large-scale spatio-temporal data set with 10 to 40 trials  pr. ferret resulting in  275 trials (films) in total. Each   trial contains  between 977 and 1720  images with time resolution equal to 0.6136 ms pr. image.  For this particular data set each  image was recorded using a hexagonal photodiode array with 464  channels and a spatial resolution  equal to 0.15 mm pr. channel (total diameter is 4.2 mm), see \cite{roland2006}. The hexagonal array is then mapped to a $25\times  25$  rectangular array to yield  one frame or image. This mapping in principle introduce some distortion of the image but we ignore that here and consider 0.15 mm pr. pixel to be the spatial resolution. 
 
Here we present an  aggregated fit obtained by first  fitting the above model to all trials  for  animal  308 (12  trials) each consisting of 977 $25\times 25$ images. Then by mean aggregating these single trial fits  we obtain  one  fit based on all trials for this animal.   Note that each of the 12 single trial fits for animal 308 are visualized in Section 3 in  the supplementary material. We have carried out this analysis   for all 13 animals -- the entire data set -- and  present the visualizations of the  remaining 12 aggregated fits in Section 2 in the supplementary material. The   estimation procedure and a snippet of the data is available from CRAN via the \textsf{R} package \verb+dynamo+, see \cite{lund2018c}.
 
For the analysis  we let  $L\coloneqq 50$  thus  allowing a   31 ms delay. For each single trial (film) the model is fitted using a  lasso regularized linear array model  derived in Section \ref{sec:plam} below. The lasso regression  is carried out for  10 penalty parameters $\lambda_1>\ldots>\lambda_{10}>0$. Here we present the fit for model 6, i.e., $\lambda=\lambda_6$, which we selected using 4-fold cross validation on the 12 trials,  see Section 1 in the supplementary material. For additional details about the specific regression setup see  \ref{subsection:regsetup}.
 
\subsection{The aggregated stimulus and network estimates}
\begin{figure}[H]
                \begin{center}
                                \includegraphics[scale=0.331]{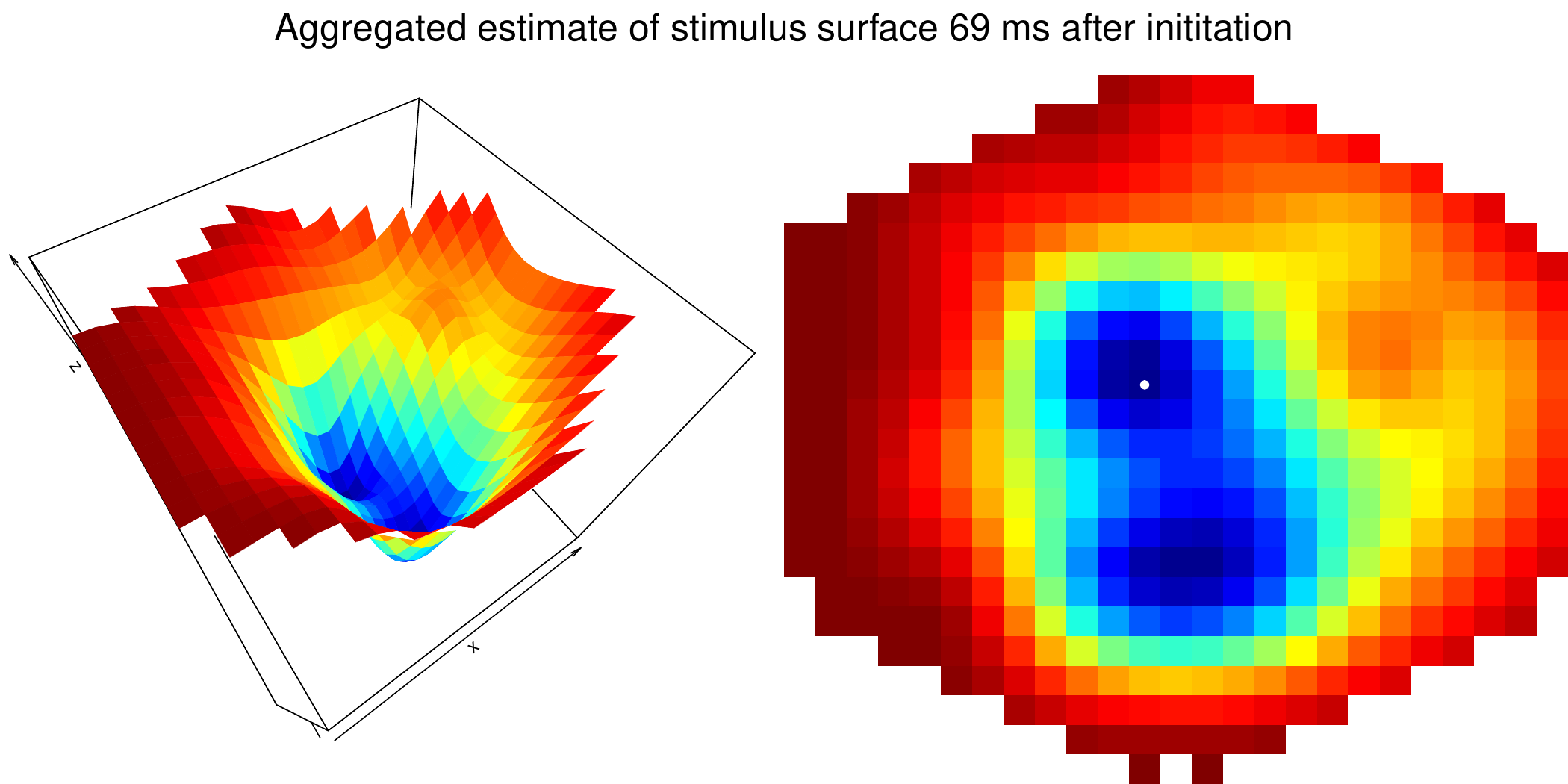}
                                \caption{Mean aggregated estimate  69 ms after stimulus onset. The white dot in the right panel indicates the  pixel visualized in Fig. \ref{fig:one}.}
                                \label{fig:two}
                \end{center}
\end{figure}

Fig.  \ref{fig:two}  shows the  estimate of the stimulus  for all pixels 69 ms after  onset.  We argue that the ``high stimulus'' areas visible in  Fig.  \ref{fig:two}  correspond to the  expected mapping of the center of field of view (CFOV), see Fig.  1 in \cite{harvey2009}.

\begin{figure}[H]
\begin{center}
\includegraphics[scale=0.331]{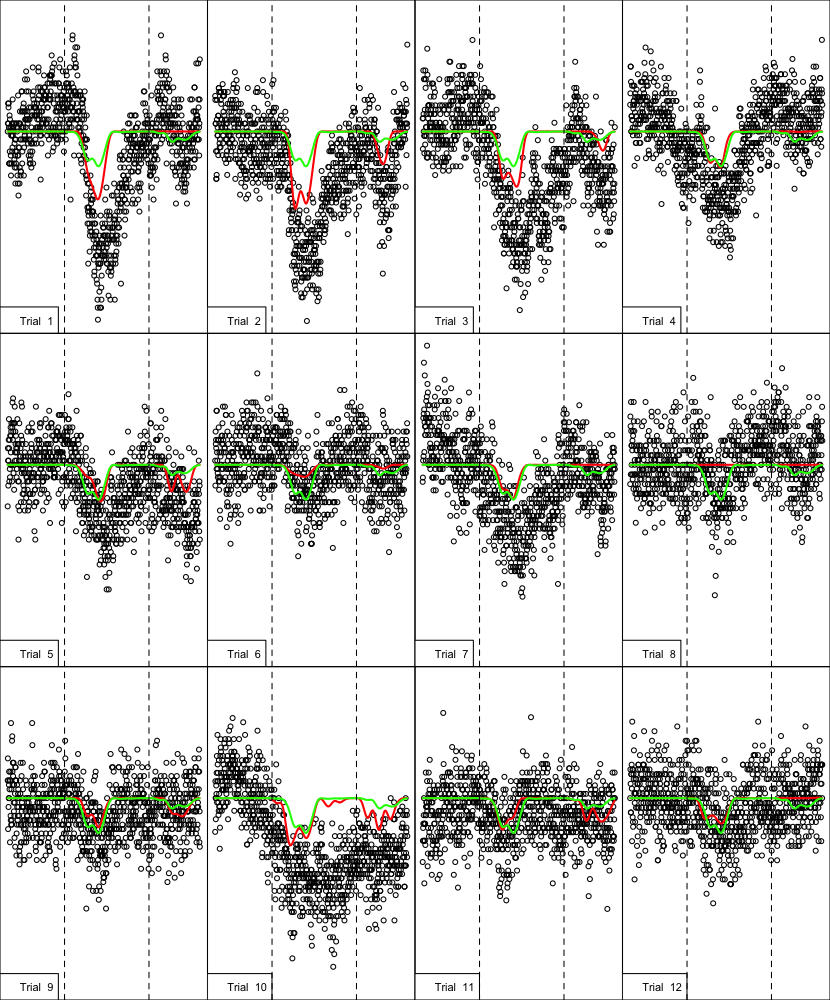}
\caption{The data observed at pixel $(x,y)$ (indicated in Fig. \ref{fig:two}) (black) and the estimate of $s(x,y,\cdot)$ for each trial (red) and the mean aggregated estimated (green). Dotted vertical lines indicate stimulus start and stop. Notice the considerable variation among the trials.}
\label{fig:one}
\end{center}
\end{figure}

For the pixel indicated with a white dot in the right panel of Fig. \ref{fig:two}, we show the raw data  for each trial   along with the trial specific estimate of  the stimulus component and the   aggregated stimulus component in Fig. \ref{fig:one}. Notice that the estimated stimulus component shows both  an on-signal after the stimulus start  and an off-signal after the stimulus stop.  Also notice the considerable variation over trials in  the raw data with some trials displaying a clear signal and others almost no signal.

Visualizing the aggregated estimate of the network is more challenging as this is quantified by the  function $w:\RR^5\to\RR$. A Shiny app   visualizing $w$ is available online (see \href{http://shiny.science.ku.dk/AL/NetworkApp/}{\texttt{shiny.science.ku.dk/AL/NetworkApp/}}), and here we present various time and space aggregated measures of propagation effects -- some of which are inspired by analogous  concepts from graph theory.
 
\begin{figure}[H]
\begin{center}
\includegraphics[scale=0.65]{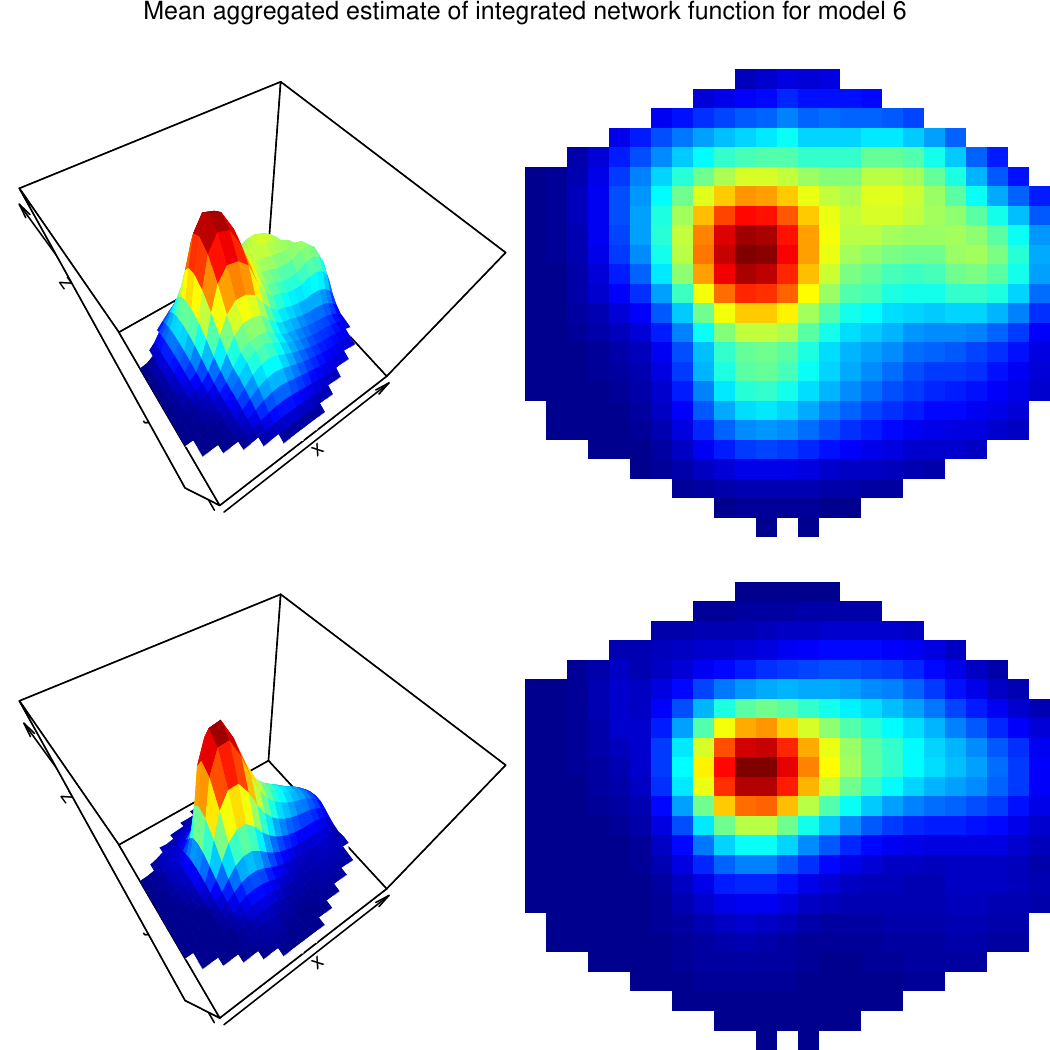}
\caption{The aggregated weight functions $\hat w^-$  (top) and $\hat w^+$  (bottom).}
\label{fig:three}
\end{center}
\end{figure}

In Fig. \ref{fig:three} below we plot the  the fitted version of the  two bivariate functions $ w^-, w^+:\RR^2\to\RR$ given by
\begin{alignat}{4}
w^-(x,y)&\coloneqq  \frac{1}{{\deg^-}(x,y)}\int_{\mathcal{S}}\int_{-\tau}^0 \vert w(x,y, x', y', t)\vert \dif{}t \dif{}x'\dif{}y',\\ 
w^{+}(x',y')&\coloneqq  \frac{1}{{\deg^+}(x',y')}\int_{\mathcal{S}}\int_{-\tau}^0 \vert w(x,y, x', y', t)\vert \dif{}t \dif{}x\dif{}y 
\end{alignat}
where
\begin{alignat*}{4}
  \deg^-(x,y)&\coloneqq \int_{\mathcal{S}}\int_{-\tau}^0 1_{ \{w(x,y, x', y', t)\neq0\}}  \dif{}t \dif{}x'\dif{}y'\quad 
  \\
  \deg^+(x',y')&\coloneqq \int_{\mathcal{S}}\int_{-\tau}^0 1_{ \{w(x,y, x', y', t)\neq0\}} \dif{}t \dif{}x\dif{}y,
\end{alignat*}
are  the aggregated  non-zero effects going in to $(x',y')$ (indegree) respectively  out  from $(x,y)$ (outdegree).
Here $ w^+(x',y')$ quantifies the effect of  $(x',y')$ on \emph{all} other coordinates relative to the  aggregated non-zero effects, that is time and space aggregated propagation effects \emph{from}  $(x',y')$.  Similarly $ w^-(x,y)$ quantifies  time and space aggregated propagation effects \emph{to}  $(x,y)$. Mean aggregated estimates of these functions are shown  in Fig. \ref{fig:three}.
 
From  bottom panel in Fig. \ref{fig:three} we see that an area is identified which across all pixels has a relatively great weight on other pixels across the 12 trials. Thus  this area is identified by   the model as important in  the  propagation of neuronal activity across  trials.  We  notice that this high output  as quantified by $ w^+$ overlaps with the strongest of the two high stimulus areas (CFOV) in Fig. \ref{fig:one}. Thus the estimated weight functions propagate primarily the direct stimulus signal.
 
From the top panel we see that the pixels receiving  propagation effects on the other hand is more scattered  around the cortex. However,  the high input areas overlap with both of  the high stimulus areas (CFOVs) in Fig. \ref{fig:one} suggesting the existence of a propagation network  connecting the high stimulus area and  the low stimulus area and  the immediate surroundings of the high stimulus  area.
 
Next for  fixed $(x',y')$  consider quantifying the aggregated (in) effects from all  points that lie $s$ spatial units away and that arrive with a delay of $t$ time units.  Letting $p$ denote the  polar coordinate parametrization   of the $s$-sphere $S_1(s)$ in $\RR^2$ with centre $(0,0)$,  
\begin{alignat*}{4}
p: [0,2\pi]\to   S_1(s),
\end{alignat*}
and  compute the desired quantity for fixed $s,t, x',y'$ as a curve integral
\begin{alignat*}{4}
\int_{S_1(s)} w(p + (x',y'),x',y',t) \dif{}p = \int_{0}^{2\pi} w\{p_1(r)+x',p_2(r)+ y',x',y',t\} \vert p'(r)\vert \dif{}r.
\end{alignat*}
Integrating this over $\mathcal{S}$ we obtain  a bivariate function
\begin{alignat*}{4}
W(s, t)\coloneqq  \int  \bigg\{ \int_{S_1(s)} w(p + (x',y'),x',y',t) \dif{}p \bigg\}\dif{}x'\dif{}y'
\end{alignat*}
giving the aggregated  effects in the entire field as a function of spatial separation $s$  (Euclidian distance)  and  temporal separation  (time delay) $t$.
From  bottom panel in Fig. \ref{fig:three} we see that an area is identified which across all pixels has a relatively great weight on other pixels across the 12 trials. Thus  this area is identified by   the model as important in  the  propagation of neuronal activity across  trials.  We  notice that this high output  as quantified by $ w^+$ overlaps with the strongest of the two high stimulus areas (CFOV) in Fig. \ref{fig:one}. Thus the estimated weight functions propagate primarily the direct stimulus signal.
 
From the top panel we see that the pixels receiving  propagation effects on the other hand is more scattered  around the cortex. However,  the high input areas overlap with both of  the high stimulus areas (CFOVs) in Fig. \ref{fig:one} suggesting the existence of a propagation network  connecting the high stimulus area and  the low stimulus area and  the immediate surroundings of the high stimulus  area.
 
\begin{figure}[H]
\begin{center}
\includegraphics[scale = 0.5]{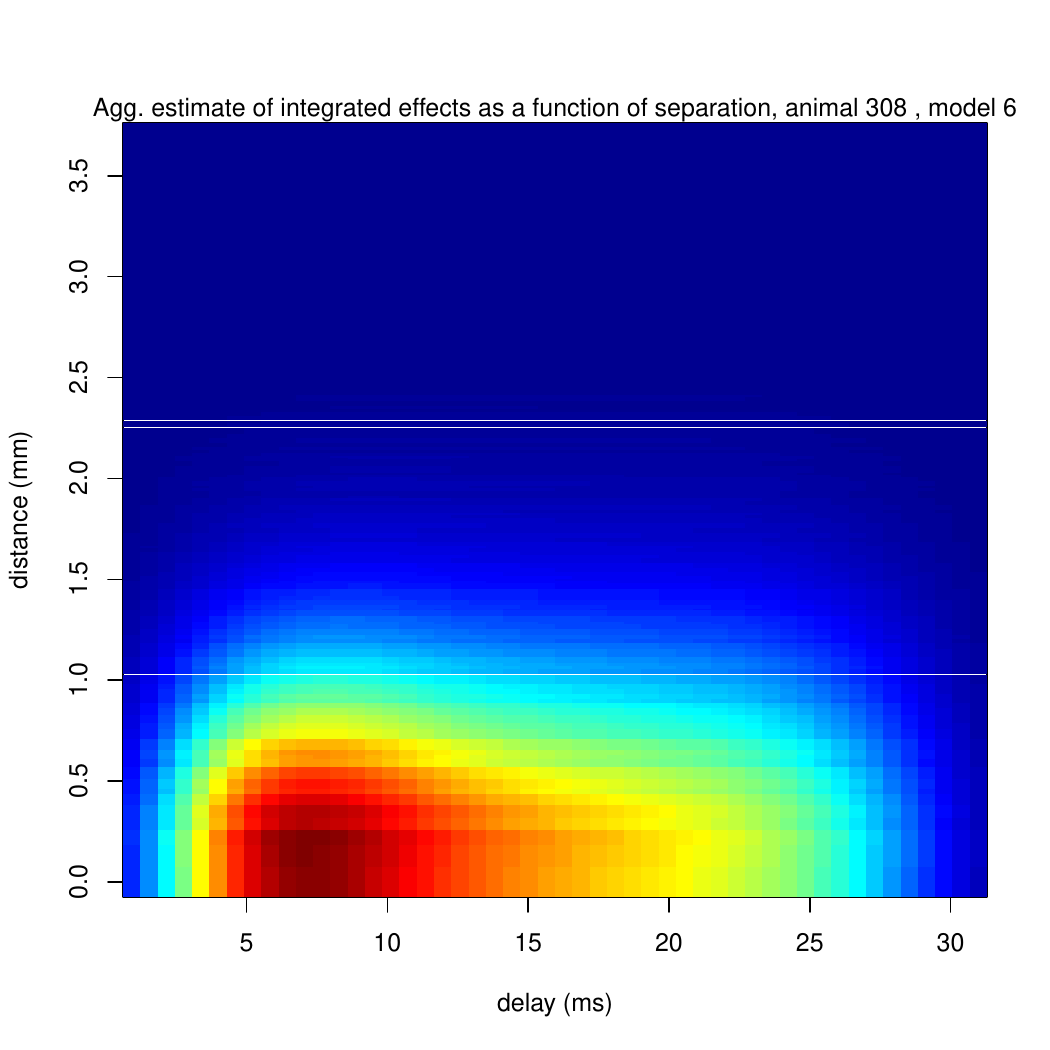}
\caption{Plot of $\hat W$ quantifying the mean aggregated estimate of propagation effects as a function of temporal and spatial separation.}
\label{fig:four}
\end{center}
\end{figure}

Fig. \ref{fig:four} summarizes the network function as a function of temporal and spatial separation. The largest effects seem to occur with a delay of around 7 ms and has   an effect on coordinates   approximately 0--0.3 mm away.    Especially the significant propagation   effects do not seem to extend beyond 1.2 mm from the source and arrive with no more than 28 ms  delay.
 
\begin{figure}[H]
\begin{center}
\includegraphics[scale = 0.45]{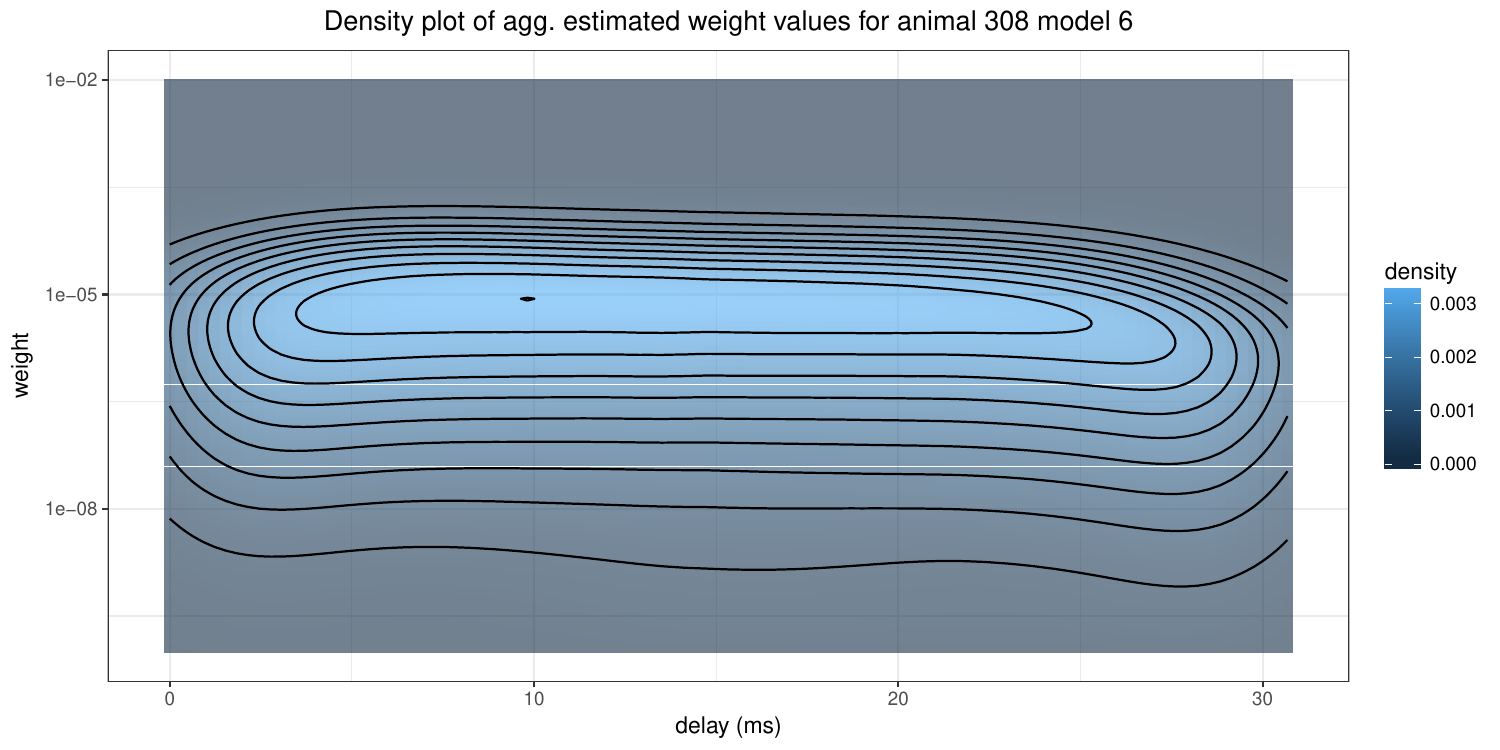}
\caption{Truncated density plots of the estimated weight values.}
\label{fig:five}
\end{center}
\end{figure}

Finally,  Fig. \ref{fig:five}  shows a  density plot of the estimated weight values in $\hat w$. The density plot is truncated as most weights are estimated to zero. From  Fig. \ref{fig:five} we can see that the  most frequent   delay  of the estimated  effects is roughly 9 to 10 ms.
 
\section{A linear   model} \label{sec:plam}
The  statistical model underlying the  inferential framework used to obtain the results in Section \ref{sec:vsd} is based on a discretized version of the SFDE \eqref{eq1}. The first step in our approach is to discretize space by aggregating the field over small areas.  Let $(\mathcal{S}_{m,n})_{m,n}\coloneqq (\mathcal{X}_m\times \mathcal{Y}_n)_{m=1,n=1}^{N_x,N_y}$ denote a partition of   $\mathcal{S}$ with $D\coloneqq N_xN_y$ elements with size ${\rm Leb}(\mathcal{S}_{m,n})=\Delta_s>0$ for all $m,n$.   We have from the integral representation of $\ar V$ in \eqref{eq7}  that
\begin{alignat}{4}\label{eq10}
\int_{\mathcal{S}_{m,n}}\ar V(&x,y,t)\dif{}x\dif{}y  =\int_0^t\bigg\{\int_{\mathcal{S}_{m,n}}s(x,y,u)\dif{}x\dif{}y\nonumber\\
  &+\int_{-\tau}^0\sum_{i,j} \int_{\mathcal{S}_{i,j}} \ar V(x',y',u+r) \int_{\mathcal{S}_{m,n}} w(x,y,x',y',r) \dif{}x\dif{}y\dif{}x'\dif{}y'\dif{}r\nonumber\\
  &+ \int_{\mathcal{S}_{m,n}}\ar V(x,y,u)h(x,y)   \dif{}x\dif{}y\bigg\}\dif{}u+\int_{\mathcal{S}_{m,n}}\int_0^t  \dif\ar  W( x, y,u) \dif{}x\dif{}y.
\end{alignat}
Here as noted in \cite{peszat1997} the last term for each $(x,y)$ is an It\^o-integral with respect to   a real valued Wiener process. Furthermore, using the covariance function for the random field from \eqref{eq4}, the  covariance of two such terms is
 \begin{alignat*}{4}
  \E\bigg\{\int_{\mathcal{S}_{m,n}}\int_0^t  \dif \ar W( x, y,u) \dif{}x\dif{}y&\int_{\mathcal{S}_{i,j}}\int_0^t  \dif \ar  W(x, y,u) \dif{}x\dif{}y'\bigg\}\\
  &=   \int_{\mathcal{S}_{m,n}}\int_{\mathcal{S}_{i,j}}   \E\{\ar W( x, y,t)  \ar  W( x', y',t) \}\dif{}x'\dif{}y'\dif{}x\dif{}y\\
&=    \int_{\mathcal{S}_{m,n}}\int_{\mathcal{S}_{i,j}}t  c (x- x',y-y')\dif{}x'\dif{}y'\dif{}x\dif{}y.
\end{alignat*}
 
We then apply  a Riemann type approximation of the space  integrals over the partition sets $\mathcal{S}_{m,n}$   on the left and the right of \eqref{eq10}. This leads us to  consider  the  $D$-dimensional real valued  stochastic process denoted $\ar{\tilde V}$, with the $(m,n)$th entry  process given by
\begin{alignat}{4}\label{eq11}
\ar { \tilde  V}_{m,n}(t) = \int_0^t\bigg\{&\ar{\tilde \Sop}_{m,n}(u)\nonumber\\
&
+\int_{-\tau}^0\sum_{i,j} \ar { \tilde V}_{i,j}(u+r) \int_{\mathcal{S}_{m,n}} w(x,y,x_i,y_j,r) \dif{}x\dif{}y\dif{}r\nonumber\\&
+   \ar{\tilde \Hop}_{m,n}\{\ar{\tilde V}(u)\}
\bigg\} \dif{}u+\int_0^t  \dif \ar { \tilde W}_{m, n}(u) \dif u,
\end{alignat}
as a space discretized model for the random field $\ar V$, where  
$\ar{\tilde \Sop}_{m,n}(u)\coloneqq \Delta_ss(x_m,y_n,u)$ and  
 $\ar{\tilde \Hop}_{m,n}\{\ar{\tilde V}(u)\}
\coloneqq \Delta_sh(x_m,y_n)\ar {\tilde V}_{m,n}(u)
$.  
 
Notice that in \eqref{eq11}  $\ar{\tilde V}(t)$ is an $N_x\times N_y$ matrix. However we might as well think of  it  as an $N_x N_y\times 1$ vector $\tilde V(t)$.  When necessary we will distinguish between the  matrix (array) form and the  vector form using the following notation.  
 
 The mapping $[i_1,\ldots,i_d]:\bigtimes_{j=1}^d\{1,\ldots, N_{i_j}\} \to \{1,\ldots, \prod_{j=1}^d N_{i_j}\}$ is a bijection, and for an array $\ar{A}\coloneqq\arr(\ar A_{i_1,\ldots,i_d})_{i_1,\ldots,i_d}$ we denote the corresponding vector as ${A}_{[i_1,\ldots,i_d]}=\ar{A}_{i_1,\ldots,i_d}$. For $d = 2$ we may take $[i_1,i_2]:=i_1+ (i_2-1)N_{i_2}$.
 
Using this notation let  $ \tilde W=\{(\tilde W_{[1,1]}(t),\ldots,\tilde W_{[N_x,N_y]}(t))\}_t$ denote   a $D$-dimen\\-sional Brownian motion on  $(\Omega, \F, \Pr)$  adapted to $(\F_t)_t$  with covariance matrix $\tilde C$ having rows given by
\begin{alignat*}{4}
\tilde C_{[m,n]}\coloneqq  (c (x_m-x_1,y_n-y_1) ,\ldots,c (x_m- x_{N_x},y_n-y_{N_y}))\Delta_s^2, 
\end{alignat*}
for each $[m,n]\in \{1,\ldots,D\}$, then  $\ar{\tilde W}$ with $\ar{\tilde W}_{i,j}=\tilde W_{[i,j]}$
is  the array version appearing in \eqref{eq11}.    
 
Aggregating  $\ar V$ over the spatial partition $ (\mathcal{S}_{m,n})_{m,n} $ leads to a multi-dimen\\-sional SDDE as an  approximate model for our data. Such models are mathematically easier to handle compared to the random field model \eqref{eq7}, see \cite{mao2007}. Especially, we can  obtain  a discrete time approximation to the solution \eqref{eq11} using an Euler scheme following \cite{buckwar2005}. The proof is in   \ref{sec:proofs}.
\begin{thm_prop} \label{prop:one}
Let $(t_k)_{k=-L}^{M}$  with $\Delta_t\coloneqq t_{k+1}-t_k>0$ denote a partition of $\mathcal{T}$. With   $(\ar{w}_\ell)_{\ell=-L}^{-1}$ a sequence of  weight matrices depending on the network function $w$, the   solution $(\tilde {\tilde V}_k)_k$ to the forward Euler scheme
\begin{alignat}{4}\label{eq12}
\tilde {\tilde V}_{k+1}&-\tilde {\tilde V}_k&=\bigg\{\tilde \Sop (t_k)+\sum_{\ell=-L}^{-1}\ar{w}_\ell\tilde {\tilde V}_{k+\ell}+\tilde \Hop(\tilde {\tilde V}_k)\bigg\}\Delta_t
+\tilde C\sqrt{\Delta_t}\epsilon_{k},
\end{alignat}
where $(\epsilon_k)_k$ are independent and  $\mathcal{N}(0,I_D)$ distributed, converges weakly to the (vectorized) solution  $\tilde V$ to \eqref{eq11}.\end{thm_prop}
Note that for each $k$, $\tilde \Sop (t_k)$ and $\tilde \Hop(\tilde {\tilde V}_k)$  are just the vectorized versions of
$\ar{\tilde \Sop }(t_k)$ and $\ar{\tilde \Hop}(\ar{\tilde {\tilde V}}_k)$ respectively with $\ar{\tilde {\tilde V}}$ the array version of ${\tilde {\tilde V}}$.
 
Next as the component functions; the stimulus functions $s$,  the network function $w$ and the short range memory $h$, are assumed to be square integrable we can use basis expansions to represent  them, i.e., for $p_s,p_w,p_h\in \NN$, 
 \begin{alignat}{4}
  s  (x,y,t)&=\sum_{q=1 }^{p_s} \alpha_q\phi_{q }(x,y,t),\label{eq16}\\
w(x,y,x',y',t)&=\sum_{q=1 }^{p_w}\beta_q  \phi_{q }(x,y,x',y',t),\label{eq17}\\
h(x,y) &=\sum_{q =1}^{p_h} \gamma_q \phi_{q }(x,y).\label{eq18}
\end{alignat}
where  $(\alpha_q)_q$, $(\beta_q)_q$ and $(\gamma_q)_q$  are  three sets of basis coefficients and $(\phi_q)_q$ is   generic notation for a  set of basis functions. We let    $p\coloneqq p_s+p_w+p_h$ denote the total number of basis functions used. Using \eqref{eq16}-\eqref{eq18}  we can formulate the dynamical model \eqref{eq12} as a vector autoregressive model. The proof is given in    \ref{sec:proofs}.
\begin{thm_prop}
There exists a $D\times 1$ vector $y_k$ and a $D\times p$ matrix $X_k$   such that we can write the dynamical model  \eqref{eq12} as an autoregression
\begin{alignat}{4}\label{eq19}
  y_k=X_k\theta+e_k, \quad k=0,\ldots,M-1
\end{alignat}
where    $(e_k)_k $ are i.i.d. $ \mathcal{N}_{D}(0, \Sigma)$ with $\Sigma\coloneqq  \tilde C^\top \tilde C\Delta_t$. Defining
\begin{alignat}{4}\label{eq412new}
y\coloneqq
\left(
\begin{array}{ccc}
  y_1   \\
    \vdots   \\
  y_M  
\end{array}
\right),\quad
X\coloneqq
\left(
\begin{array}{ccc}
  X_1   \\
    \vdots   \\
  X_M  
\end{array}
\right),
\end{alignat}  
  the associated  negative  log-likelihood can be written as
  \begin{alignat}{4}\label{eq412}
  \ell_X(\theta, \Sigma)= \frac{M}{2}\ln \vert \Sigma\vert+\big\Vert \big(I_M\otimes \Sigma^{-1/2}\big)(y  - X\theta)\big\Vert _2^2,
\end{alignat}
with $I_M$ the $M\times M$ identity matrix.
\end{thm_prop}
 
 We note that due to the structure of the linear model \eqref{eq19} the MLEs of $\theta $ and $\Sigma$ are not available in closed form. Especially the  normal equations characterizing  the MLEs are coupled in this model leading to a generalized least squares type estimation problem.  Furthermore, the $DM\times p$ design matrix $X$ will for realistically sized data become very large making it infeasible to  fit the model \eqref{eq19} directly by minimizing \eqref{eq412} .
 Next we will discuss how to compute regularized estimates of the parameters $\theta$ and $\Sigma$ by exploiting the array structure of the problem to reformulate the autoregressive model \eqref{eq19} as a (partial) linear array model. This in turn makes the computations involved in the fitting procedure feasible.
  
  \subsection{Penalized linear array model}
In order to obtain  time and space localized estimates of the  component functions,  we will   minimize a regularized version of  \eqref{eq412}. Letting $\Omega \coloneqq  \Sigma^{-1}$ denote the precision matrix, this is achieved  by solving the    unconstrained  problem
\begin{alignat}{4}\label{eq24}
\min_{\theta\in \RR^{p},\Omega \in M^{D\times D}} \ell_X(\theta, \Omega^{-1}) + \lambda  J_1(\theta) + \nu J_2(\Omega),
\end{alignat}
where $J_1$ and $J_2$ are convex penalty functions and $\lambda\geq0  $ and $\nu\geq0  $ the penalty parameters  controlling the amount of regularization.  For    non-differentiable penalty functions $J_i$, $i=1,2$,  solving \eqref{eq24}  results in  sparse estimates of  $\theta$ and $\Omega$.  In  the following we will use  the lasso or $\ell_1$  penalty, i.e.,   $J_1=J_2=\Vert\cdot\Vert_1$, see \cite{tibshirani1996}.
 
Following \cite{rothman2010} we  solve \eqref{eq24} using their approximate MRCE algorithm.   In our setup the steps are  as follows:
 
\begin{enumerate}
\item For fixed  $\hat \Omega $ and  each penalty parameter $\lambda_1\geq\ldots\geq\lambda_K$, $K\in \NN$, solve
\begin{alignat}{4}\label{eq414}
\min_{\theta\in \RR^{p}} \ell_X(\theta,{\hat \Omega}^{-1}) + \lambda_i            J_1(\theta).
\end{alignat}
\item For $i\in \{1,\ldots,K\}$  let $\hat \theta_{\lambda_i}$ denote the estimate from step 1,  let $\hat\Sigma_{R,i}\coloneqq (y-X\hat \theta_{\lambda_i})^\top(y-X\hat \theta_{\lambda_i})$  and  use  graphical lasso, see \cite{friedman2008}, to solve
\begin{alignat*}{4}
\hat \Omega\coloneqq  \arg\min_{\Omega} \tr(\hat\Sigma_{R,i}\Omega)-\ln\vert \Omega\vert+\nu J_2(\Omega).
\end{alignat*}
\item Repeat step 1 and 2 with weighted data $\tilde y\coloneqq \hat \Omega^{1/2} y$ and $\tilde X \coloneqq  \hat \Omega^{1/2} X$.
\end{enumerate}
 
Now, solving the problem \eqref{eq414} requires a numerical procedure. However   from a  computational viewpoint evaluating $\ell_X$ given in \eqref{eq412} becomes infeasible as  the design matrix $X$  in practice is enormous.  For instance for each of the  trials  considered in Section \ref{sec:vsd}  and the setup described in \ref{subsection:regsetup} the design $X$  takes up around $200$ GB of memory. In addition, even if we could allocate the memory needed to store $X$ the time needed to compute the entries in $X$ would be considerable and any algebraic operation involving $X$ potentially computationally infeasible.
 
The solution to this computational problem is to choose basis functions in the representations \eqref{eq16}-\eqref{eq18} that will lead to a decomposable design matrix $X$.  In particular,    assume  $p_x,p_y,p_t,p_\ell \in\NN$ such that   $p_s=p_xp_yp_t$, $p_w=p_xp_yp_xp_yp_\ell $ and $p_h=p_xp_y$. Then using  tensor product basis functions we can write \eqref{eq16}-\eqref{eq18} as
  \begin{alignat}{4}
  s  (x,y,t)&=
  \sum_{j_1=1 }^{p_x}
\sum_{j_2 =1}^{p_y}
\sum_{j_3 =1}^{p_t}
\ar{   \alpha}_{j_1,j_2,j_3}\phi^x_{j_1}(x)\phi^y_{j_2}(y)\phi^{t}_{j_3}(t), \label{eq21new} \\
w(x,y,x',y',t)&=
\sum_{j_1=1 }^{p_x}
\sum_{j_2=1 }^{p_y}
\sum_{j_3 =1}^{p_x}
\sum_{j_4 =1}^{p_y}
\sum_{j_5=1 }^{p_\ell }
\ar{\beta}_{j_1,j_2,j_3,j_4,j_5} \phi^x_{j_1}(x)
\phi^y_{j_2}(y)\phi^x_{j_3}(x')\phi^y_{j_4}(y')\phi^{l}_{j_5}(t), \label{eq22new}\\
h(x,y)&=
\sum_{j_1=1} ^{p_x}
\sum_{j_2=1} ^{p_y}
\ar{\gamma}_{ j_1,j_2} \phi^x_{j_1}(x)\phi^y_{j_2}(y).  \label{eq23new}
\end{alignat}
Note that $\ar{\alpha},\ar{\beta}$ and $\ar{\gamma}$ are the array versions of $\alpha$, $\beta$ and $\gamma$ respectively.
 
Using \eqref{eq21new}-\eqref{eq23new}, it turns out that  $X$ can essentially be componentwise tensor factorized, see the proof of Proposition \ref{prop:three} in  \ref{sec:proofs}, implying that we can   perform algebraic operations involving $X$ without having to construct it. This is achieved by using the   so called rotated $H$-transform from \cite{currie2006}.
     \begin{thm_def}\label{def:two}
  Let  $\ar A$ be  $p_1\times \cdots\times p_d$  and $X_i$  $n_i\times p_i$  for $i\in \{1,\ldots,d\}$. The rotated $H$-transform is defined as the map  $\rho$  such that
\begin{alignat}{4}\label{eq22}
X_d\otimes \cdots\otimes X_1 \ve(\ar A)= \ve[\rho\{X_d,\rho(\ldots,\rho[X_1, \ar A])\} ],
\end{alignat}
where $\ve$ is the vectorization operator. 
\end{thm_def}
 
The above definition is not very enlightening in terms of how $\rho$ actually computes the matrix-vector product. These details can be found in \cite{currie2006}. Definition \ref{def:two} shows, however, that for a tensor structured matrix  we can compute matrix-vector products via $\rho$  using only its tensor components. This makes $\rho$  very memory efficient. Furthermore, as discussed in \cite{deboor1979}, \cite{buis1996}, \cite{currie2006}, $\rho$ is also computationally efficient as the left hand side of \eqref{eq22} takes  more multiplications to compute than the right hand side. In this light the following  proposition is relevant. The proof is in  \ref{sec:proofs}.
 
\begin{thm_prop} \label{prop:three}
There exists arrays $\phi^{x},\phi^{y},\phi^{t},\Phi^{xy\ell},\Phi^{x},\Phi^{y}$ and  $\ar{C}$ such that
\begin{alignat}{4}
X\theta=\ve\Big[ &\rho[\phi^{t}, \rho\{\phi^{y}, \rho(\phi^{x}, \ar{\alpha})\}]
&+\rho[\Phi^{xy\ell}, \rho\{\Phi^{y}, \rho(\Phi^{x},  \ar{\beta})\}] + \ar {\tilde{\tilde V}}_{-1} \odot \ar C\Big].\label{eq4.15}
\end{alignat}
   Here $\ar {\tilde{\tilde V}}_{-1} \coloneqq (\ar {\tilde{\tilde V}}_k)_{k=-1}^{M-1}$, $\phi^{\mathrm{x}}$ and $\Phi^\mathrm{x}$ are  $N_\mathrm{x}\times p_\mathrm{x}$ for $\mathrm x\in\{t,x,y\}$, $\Phi^{xy\ell}$  is $ M \times p_x p_y p_\ell $,  $\ar C$ is $N_x\times N_y\times M$ and $\odot$ denotes the Hadamard product.
\end{thm_prop}
 
Proposition \ref{prop:three} shows that we can in fact write  the linear model \eqref{eq19} as a three component partial 3-dimensional linear array model, see \cite{lund2017} for multi-component array models. Note we say partial since the last component does not have a full tensor decomposition.
This in turn has important computational consequences for the model fitting procedure  as shown in \cite{lund2017}, where a  gradient descent proximal gradient (gd-pg) algorithm is proposed for this kind of model setup. The gd-pg  algorithm uses a minimal amount of memory and can exploit  efficient array arithmetic (the $\rho$ operator in \eqref{eq4.15})  while  solving the non-differentiable penalized estimation problem \eqref{eq414}.  Especially it is fairly straightforward to construct an estimation procedure based on the gd-pg algorithm that will solve the penalized problem \eqref{eq414}  for all three model components at once, that is  minimize the penalized log-likelihood over all parameters $\alpha,\beta, \gamma$ simultaneously.
 
\subsection{Block relaxation model fitting}
We propose a variation of the gd-pg algorithm, which uses a block relaxation scheme, to solve \eqref{eq414} for one parameter block (e.g., $\ar \alpha$) at a time while fixing the rest ($\ar \beta$ and $\ar \gamma$), see \cite{de1994}. We note that due to the additive mean structure this  approach corresponds to a  back-fitting algorithm  where the  partial residuals, resulting from fixing all but one parameter block,  are fitted to data for one  model component at the time.
 
The block relaxation approach is directly motivated by the application to neural field models as a way to achieve a particular structure on the estimated stimulus component. Without any structural constraints, the stimulus component would effectively just be a smoothing of the observed signal, and we would not achieve the decomposition of the drift into a direct stimulus component and a propagation component. We argue that the temporal evolution in the direct reaction to the stimulus is homogeneous across space, though the size of the direct stimulus component vary. Thus the stimulus component consists of a spatially modulated temporal signal.
 
This model constraint corresponds to assuming that  $s$ can be factorized into a product of a bivariate function of  space and a univariate function of time. Representing each of these factors in a tensor basis then leads to the representation
\begin{alignat*}{4}
  s(x,y,t) =\sum_{k=1}^{p_t} \sum_{j=1}^{p_y}\sum_{i=1}^{p_x}\phi^t_k(t) \phi^y_j(y)\phi^x_i(x)\zeta_k\ar\eta_{i,j}.
\end{alignat*}
 
Using a block relaxation scheme we  can place the reduced rank   restriction
  \begin{alignat*}{4}
  \ar\alpha_{i,j,k}=\zeta_k\ar\eta_{i,j}
\end{alignat*}
on the  coefficients in the  3-dimensional coefficient array $ \ar\alpha$ when solving the  subproblem pertaining to the stimulus component. We note that the resulting reduced rank subproblem may then be solved with a modified version of the gd-pg algorithm where  (again) a block relaxation algorithm  is incorporated to obtain the desired factorization of the parameter array. 
 
Applying  this block relaxation approach we may use the existing  software in the \verb+glamlasso+  \textsf{R} package (see \cite{lund2018}) to obtain the estimates for the parameter blocks $\ar \alpha,\ar \beta,\ar \gamma$ respectively.  In   \ref{sec:impl} below we give additional details on how to construct the algorithm, which is implemented in the \textsf{R} package \verb+dynamo+ available on CRAN, see \cite{lund2018c}.
 
\section{Discussion}\label{sec:conc}
The VSD imaging data analyzed in this paper exemplifies noisy spatio-temporal  array   data with  a potentially complicated spatio-temporal dependence structure. Our proposed methodology entails  modelling this type of data as a non-stationary stochastic process (random field) in order to obtain a   viable statistical model.   We note that one immediate   challenge inherent in this methodology is to find a sensible decomposition of the drift into a stimulus component and a propagation component such that the statistical model can yield meaningful estimates of these components. Two related issues turned out to be of particular importance: first, the resulting fitted dynamical model should be   stable;  and second,  both the fitted stimulus component and the fitted propagation component  should  be non-zero. The reduced rank procedure we used for fitting the stimulus component specifically addressed the latter issue while the sparsity inducing penalty addressed the former.
 
Next we  showed how to   exploit the particular array-tensor  structure inherent in this specific data-model  combination, to obtain a computationally tractable estimation procedure.  The computational challenge was primarily addressed via the discretization schemes and basis expansions  which resulted in a linear array model, that can be fitted using the algorithm  proposed in \cite{lund2017}. This allowed us to obtain a design matrix free  procedure with a minimal memory footprint that  utilizes  the highly efficient array arithmetic, see \cite{deboor1979}, \cite{buis1996} and \cite{currie2006}. Consequently we were able to fit the model to VSD imaging data  for a single trial (film) with  625 pixels in the  spatial dimension and recorded over thousands of time points while modeling  very long time delays,  on a standard laptop computer in a matter of minutes.
 Given the results in \cite{lund2017} we expect our procedure to scale  well with the size of the data.
 
In conclusion we highlight that
both in terms of interpretability and computability  the methodology  developed in this paper  compared to existing methods  is particularly attractive for lager-scale applications.
For instance we note that   fitting data on  this  scale  is computationally prohibitive using conventional time series techniques. Unrestricted vector autoregressions (VAR) has a parameter dimension that grows quadratically in the size of the spatial dimension, see \cite{fan2011}, and would be difficult to interpret and suffer from large variances on the parameter estimates. See, e.g., \cite{valdes2005}  for a VAR model  applied to fMRI brain image  or the approach in \cite{davis2016} to  high dimensional  VAR analysis, both considering data with size on a much smaller scale  than the VSD imaging data. Specifically, fitting a traditional VAR type model to the data and setup considered here would result in a model with  $LD^2=19,531,250$ parameters.  In comparison our proposed estimation framework only uses 46,848 parameters of which only around 1,800 are non-zero in the aggregated estimate  (model no. 6, animal 308).  Within our methodological framework, this dimension reduction is obtained by modelling the  array data as a random field, which makes it possible to represent the drift components in terms of  smooth functions that in turn are easier to interpret compared to millions of single parameter estimates. In addition, by using a combination of local basis functions and sparsity inducing penalties we achieve even more regularization  without restricting the model to a narrow parametric class. 
 \section*{Acknowledgments}
The research was supported by VILLUM FONDEN via research grant 13358. 

\appendix
\section{Proofs}\label{sec:proofs}
\begin{proof}[\bfseries Proof of Proposition 1]
With $w$ the network function  let $\tilde w $ denote  a  $D\times D$ matrix-valued signed  measure on $[-\tau,0]$   with  density        $\tilde F:\RR\to\RR^{D\times D}$, 
\begin{alignat*}{4}
\tilde F(t)\coloneqq
\left(
\begin{array}{ccc}
\int_{\mathcal{S}_{1,1}} w(x,y,x_i,y_j,t) \dif{}x\dif{}y&   \ldots&\int_{\mathcal{S}_{N_x,N_y}} w(x,y,x_i,y_j,t) \dif{}x\dif{}y  \\
  \vdots&   \ddots&\vdots   \\
  \int_{\mathcal{S}_{1,1}} w(x,y,x_{N_x},y_{N_y},t) \dif{}x\dif{}y& \ldots  & \int_{\mathcal{S}_{N_x,N_y}} w(x,y,x_{N_x},y_{N_y},t) \dif{}x\dif{}y \\
\end{array}
\right),
\end{alignat*}
with respect to the  Lebesgue measure on $\RR$.   Following \cite{buckwar2005} we  consider a    stochastic delay differential equation for a $D$-dimensional (vector) process $\tilde V$ given by
\begin{alignat}{4}
\dif \tilde V(t)=\bigg[\tilde \Sop(t)+ \int_{-\tau}^0\tilde w(\dif r) \tilde V(t+ r)+\tilde \Hop\{\tilde V(t)\}\bigg]\dif t +\tilde C\dif \tilde W(  t)\label{eq13}\\
\tilde V(0)\in \RR^D,\quad  \tilde V(u)=\tilde V_{0}(u)\quad u\in (-\tau,0)\label{eq14},
\end{alignat}
where $\tilde V(0)$ and  $\tilde V_0 \in C([-\tau, 0], \RR^{D} )$ are initial conditions. A solution to \eqref{eq13}   and \eqref{eq14} is then given by the integral equation  \eqref{eq11} giving the coordinate-wise  evolution in the space discretized model for the random field  $\ar V$.
 
The  sequence of $D\times D$ weight matrices  $ (\ar{w}_\ell)_\ell$ is   defined by
\begin{alignat*}{4}
\ar{w}_\ell \coloneqq \int_{-\tau}^0\boldsymbol{1}_{[t_{\ell},t_{\ell+1})}(s)\dif \tilde w(s)=\int_{-\tau}^0\boldsymbol{1}_{[t_{\ell},t_{\ell+1})}(s)\tilde F(s)\dif s,
\end{alignat*}
where   $\boldsymbol{1}_{[t_{\ell},t_{\ell+1})}$ is equal to $I_{D}$ on  $[t_{\ell},t_{\ell+1})$ and zero otherwise. Especially the entry in the $[m,n]$th row and $[i,j]$th column of   $\ar{w}_\ell$ is given as 
\begin{alignat}{4}\label{eq15}
\ar{w}_{[m,n],[i,j],\ell}=\int_{t_{k}}^{t_{k+1}}\int_{\mathcal{S}_{m,n}} w(x,y,x_{i},y_{j},s) \dif{}x\dif{}y
\dif s,
\end{alignat}
and we note that $\ar{w}$ is a $ D \times D \times  L$ array.
Letting $(\epsilon_k)_k$ denote a sequence of i.i.d. $\mathcal{N}(0,I_{D})$ variables the  Euler scheme from \cite{buckwar2005} is now given by the $D$-dimensional discrete time (vector) process $\tilde {\tilde V}=(\tilde {\tilde V}_k)_k$ solving the stochastic difference equation  \eqref{eq12} for $k \in \{ 0,\ldots, M -1\}$  with initial  conditions $\tilde{\tilde V}_\ell =\tilde V_0(t_\ell)$ for $\ell \in \{-L,\ldots,-1\}$ and $\tilde{\tilde V}_0 =\tilde V(0)$. Thus by Theorem 1.2 in \cite{buckwar2005} and using that the deterministic function $s$ is continuous, for $\Delta_t\downarrow 0$,    $\tilde {\tilde V}$ defined in \eqref{eq12}    converges weakly to the  (vector) process solving  SDDE \eqref{eq13} and \eqref{eq14},  i.e., the vector process $\tilde V$ with   evolution identical to that of $\ar{\tilde V}$ given by   \eqref{eq11}.
\end{proof}
Note $\tilde{\tilde V}$ is simply the vectorized version of $\ar{\tilde{\tilde V}}$, that is $\tilde{\tilde V}_{[i,j],k}=\ar{\tilde{\tilde V}}_{i,j,k}$.
\begin{proof}[\bfseries Proof of Proposition 2]
First using the expansion in \eqref{eq17} we can write the entries in  each weight matrix $\ar{w}_\ell$ from \eqref{eq15} as
\begin{alignat*}{4}
\ar{w}_{[m,n],[i,j],\ell}
&=   \sum_{q }\beta_q \int_{t_{\ell}}^{t_{\ell+1}}\int_{\mathcal{S}_{m,n}} \phi_{q }(x,y,x_i,y_j,s) \dif{}x\dif{}y
  \dif s.
\end{alignat*}
Then we can write the $[m,n]$th coordinate  of the vector process from  \eqref{eq12} as
\begin{alignat}{4}
\tilde {\tilde V}_{[m,n],k+1}
&=\bigg\{ \tilde s (x_m,y_n,t_k)+\sum_{\ell=-L}^{-1}
\sum_{i,j}
\ar{w}_{[m,n],[i,j],\ell} \tilde {\tilde V}_{[i,j],{k+\ell}}
\nonumber\\&\phantom{=\bigg\{}
+
(1+\tilde h(x_m,y_n))
\tilde {\tilde V}_{[m,n],k}\bigg\}\Delta_t+\tilde C_{[m,n]}\sqrt{\Delta_t}\epsilon_{k}
\nonumber\\
&=  \Delta_t\sum_{q} \alpha_q\phi_{q}(x_m,y_n,t_k)+ \beta_q  \arr{F}_{[m,n],k,q}+ \gamma_q\phi_{q}(x_m,y_n)\tilde {\tilde V}_{[m,n],k} \nonumber\\&\phantom{=\bigg(}
  +\tilde C_{[m,n]}\sqrt{\Delta_t}\epsilon_{k}\label{eqa4}
\end{alignat}
where
\begin{alignat}{4}\label{eq26}
  \arr{F}_{[m,n], k,q}&\coloneqq   \sum_{l=-L}^{-1}\sum_{i,j}\tilde {\tilde V}_{[i,j],k+l}\int_{t_{l}}^{t_{l+1}}\int_{\mathcal{S}_{m,n}} \phi_{q}(x,y,x_i,y_j,r) \dif{}x\dif{}y  \dif r
\end{alignat}
for $q \in \{1,\ldots,p_w\}$. Then letting $\theta\coloneqq(\alpha,\beta,\gamma)^\top$,
$y_k\coloneqq \tilde {\tilde V}_{k+1}$, 
$e_k\coloneqq \tilde C_{[m,n]}\sqrt{\Delta_t}\epsilon_{k}$, 
and
$X_k\coloneqq (    {S}_k\mid     {F}_k\mid     {H}_k)$
with
\begin{alignat}{4}
   {S}_k&\coloneqq  \left(
\begin{array}{ccc}
\phi_{1}(x_1,y_1,t_k) &\cdots&\phi_{p_s}(x_1,y_1,t_k)
\\
\vdots&\vdots&\vdots\\
\phi_{1}(x_{N_x},y_{N_y},t_k) &\cdots&\phi_{p_s}(x_{N_x},y_{N_y},t_k)
\end{array}
\right),\quad
    {F}_k\coloneqq \left(
\begin{array}{ccc}
  \arr{ F}_{[1,1],k}
\\
\vdots\\
\arr{  F}_{[N_x,N_y],k}
\end{array}
\right),
\nonumber \\
     {H}_k &\coloneqq  \left(
\begin{array}{ccc}
\phi_{1}(x_1,y_1)\tilde {\tilde V}_{[1,1],k}&\cdots&\phi_{p_h}(x_1,y_1) \tilde {\tilde V}_{[1,1],k}\\
\vdots&\vdots&\vdots\\
\phi_{1}(x_{N_x},y_{N_y}) \tilde {\tilde V}_{[N_x,N_y],k}&\cdots&\phi_{p_h}(x_{N_x},y_{N_y}) \tilde {\tilde V}_{[N_x,N_y],k}
\end{array}
\right).\label{eq31}
\end{alignat}
the model equation \eqref{eq19} follows from \eqref{eqa4}.
 
To obtain the likelihood \eqref{eq412}  note that  the  transition density for the model is
\begin{alignat*}{4}
f(y_k\mid X_k) = (\sqrt{2\pi})^{-D} \vert \Sigma\vert^{-1/2}\exp\bigg\{-\frac{1}{2} (y_k-X_k\theta)^\top \Sigma^{-1}(y_k-X_k\theta) \bigg\}.
\end{alignat*}
As $y_k\mid X_k$ is independent of  $y_0,\ldots y_{k-1},X_0,\ldots X_{k-1}$
we get  by successive conditioning that we can write the joint conditional density as
\begin{alignat*}{4}
f(y_0,\ldots&,y_{M-1}\mid X_0,\ldots,X_{M-1})  \\
&= (\sqrt{2\pi})^{-MD} \vert \Sigma\vert^{-M/2}\exp\bigg\{-\frac{1}{2} \sum_{k=0}^{M-1}(y_k-X_k\theta)^\top \Sigma^{-1}(y_k-X_k\theta) \bigg\}.
\end{alignat*}
Taking $-\ln$ yields the negative log-likelihood
\begin{alignat*}{4}
\ell_X(\theta,\Sigma)&  \coloneqq  \frac{M}{2}\ln \vert \Sigma\vert+\frac{1}{2} \sum_{k=0}^{M-1}(y_k-X_k\theta)^\top \Sigma^{-1}(y_k-X_k\theta).
\end{alignat*}
With  $X$ and $y$ as in \eqref{eq412new} it follows that 
\begin{alignat*}{4}
\big(I_M\otimes \Sigma^{-1/2}\big)(y- X\theta)&=
\left(
\begin{array}{ccc}
\Sigma^{-1/2}  (y_0-X_0\theta)   \\
    \vdots   \\
\Sigma^{-1/2}  (y_{M-1}-X_{M-1}\theta)
\end{array}
\right)\\
&=
\left(
\begin{array}{ccc}
\sum_j^D\Sigma^{-1/2}_{1,j}  (y_0-X_0\theta)_j   \\
    \vdots   \\
\sum_j^D\Sigma^{-1/2}_{D,j}  (y_0-X_0\theta)_j   \\
\vdots\\
\sum_j^D\Sigma^{-1/2}_{D,j}  (y_{M-1}-X_{M-1}\theta)_j   \\
\end{array}
\right).
\end{alignat*}
Hence,
\begin{alignat*}{4}
\big\Vert\big( I_M\otimes \Sigma^{-1/2}\big)(y- X\theta)\big\Vert _2^2
&
=\sum_{k=0}^{M-1}\sum_i^D\bigg\{\sum_j^D\Sigma^{-1/2}_{i,j}  (y_k-X_k\theta)_j \bigg\}^2\\ &
= \sum_{k=0}^{M-1}\Vert\Sigma^{-1/2}(y_k-X_k\theta)\Vert_2^2 \\&
=\sum_{k=0}^{M-1}(y_k-X_k\theta)^\top \Sigma^{-1}(y_k-X_k\theta)
 \end{alignat*}
yielding the expression for the negative log-likelihood given in \eqref{eq412}.
\end{proof}
 
\begin{proof}[\bfseries Proof of Proposition 3]
Noting that 
\begin{alignat*}{4}
  ( {S}\mid  {F}\mid  {H})\coloneqq
  \left(
\begin{array}{ccc}
   {S}_{1}& {F}_1& {H}_1   \\
  \vdots &\vdots&\vdots  \\
   {S}_{M}& {F}_M& {H}_M   \\
\end{array}
\right)=X
\end{alignat*}
  the claim follows if we show that  $ {S}$ and $ {F}$ are appropriate 3-tensor matrices and that $H\gamma=
\ve(  \ar{\tilde {\tilde V}}_{-1}\odot \ar C)$ for an appropriately defined array $\ar C$.
 
Letting $\phi^x\coloneqq \{\phi_q(x_i)\}_{i,q}$ denote a  $N_x\times p_x$ matrix with  $p_x$ basis functions evaluated at $N_x$ points in the $x$ domain it follows directly from the definition of the tensor product  that we can write  $S=\phi^x\otimes\phi^y\otimes\phi^t$.
 
Next let $\Phi^x\coloneqq \{\int_{\mathcal{X}_i}\phi_q(x)\}_{i,q}$ denote  the integrated version of $\phi^x$. Then inserting  the tensor basis functions  in to \eqref{eq26} we can write
\begin{alignat*}{4}
&\arr{F}_{[m,n],k,[q_1,q_2,q_3,q_4,q_5]}
\\&\phantom{.........}=\sum_{\ell=-L}^{-1}\sum_{i,j}\tilde {\tilde V}_{[i,j],k+\ell}
\int_{t_{l}}^{t_{\ell+1}}\int_{\mathcal{S}_{m,n}}
\phi_{q_1}(x)
\phi_{q_2}(y)
\phi_{q_3}(x_i)
\phi_{q_4}(y_j)
\phi_{q_5}(s) \dif{}x\dif{}y  \dif s\\
&\phantom{.........}=
\int_{\mathcal{X}_m}\phi_{q_1}(x)\dif{}x
\int_{\mathcal{Y}_{n}}\phi_{q_2}(y) \dif{}y 
\sum_{\ell=-L}^{-1}\sum_{i,j}\tilde{\tilde V}_{[i,j],k+\ell}
\phi_{q_3}(x_i)
\phi_{q_4}(y_j)
\int_{t_{\ell}}^{t_{\ell+1}}
\phi_{q_5}(s)\dif s.
\end{alignat*}
Letting $\Phi^{xy\ell}$ denote a $M\times p_xp_yp_\ell$ matrix  with entries
\begin{alignat}{4}\label{eqa5}
\Phi_{k,[q_3,q_4,q_5]}^{xy\ell}\coloneqq \sum_{\ell=-L}^{-1}\sum_{i,j}\tilde{\tilde V}_{[i,j],k+\ell}
\phi_{q_3}(x_i)
\phi_{q_4}(y_j)
\int_{t_{\ell}}^{t_{\ell+1}}
\phi_{q_5}(s)\dif s 
\end{alignat}
we can write   $F=\Phi^{xy\ell} \otimes\Phi^{y}\otimes\Phi^{x}$.
 
Finally let  $\ar C$ be a $N_x\times N_y\times M$ array such that  $\ar C_k=\phi^{y}\ar{\gamma} (\phi^{x})^\top$.
With $H_k$ given in \eqref{eq31} we can write
\begin{alignat*}{4}
  H_k\gamma&= \left( \begin{array}{ccc}
  \tilde {\tilde V}_{[1,1],k} \sum_{q_1,q_2} \phi_{q_1}^x(x_1)\phi_{q_2}^y(y_1)\gamma_{[q_1,q_2] }\\
\vdots\\ 
\tilde{\tilde   V}_{[N_x,N_y],k} \sum_{q_1,q_2} \phi_{q_1}^x(x_{N_x})\phi_{q_2}^y(y_{N_y})\gamma_{[q_1,q_2] }\\
  \end{array}
\right)\\
&=\tilde{\tilde  V}_k \odot \phi^x\otimes \phi^y\gamma\\
&=\ve(\ar{\tilde{\tilde  V}}_k \odot \ar C_k)
\end{alignat*}
showing that    $H\gamma =   \ve (\ar{\tilde{\tilde V}}_{-1}\odot \ar C)$.
\end{proof}
 
\section{Tensor product basis}
\label{sec:tensor}
Consider a  $d$-variate function   $f\in L^2(\RR^d)$ that we want to represent using  some basis expansion. Instead of directly specifying a basis for   $L^2(\RR^d)$ we will  specify $d$ marginal sets of  univariate functions
 \begin{alignat}{4}\label{three}
( \phi_{1,k_1})_{k_1=1}^{\infty},\ldots,( \phi_{d,k_d})_{k_d=1}^{\infty}
\end{alignat}
   with each marginal set  a basis for  $L^2(\RR)$.  Then for any $(k_1,\dots,k_d)\in \NN^d$ we may define a $d$-variate function   $\pi_{k_1,\ldots,k_d}:\RR^d\to \RR$  via the product of the marginal functions, i.e.,
\begin{alignat}{4}\label{four}
(x_1,\ldots,x_d)\mapsto \pi_{k_1,\ldots,k_d}(x_1,\ldots,x_d)\coloneqq  \prod_{i=1}^d \phi_{i,k_i}(x_i).
\end{alignat}
Since each  marginal set of functions in \eqref{three}  constitutes a basis of $L^2(\RR)$ it follows  using Fubini's theorem, that  the  induced set of  $d$-variate functions  \\$( \pi_{k_1,\ldots,k_d})_{k_1,\ldots,k_d}$   constitutes  a basis of    $L^2(\RR^d)$. Especially again by Fubini's theorem we note that if the functions in \eqref{three} all are orthonormal marginal bases then they generate an orthonormal  basis of  $L^2(\RR^d)$. Finally, if the  marginal functions have compact support then the induced $d$-variate  set of functions will also have compact support.
 
\section{Implementation details}\label{sec:impl}
Here we    elaborate a little on various   details pertaining to the   implementation of the inferential procedure for the specific data considered in Section \ref{sec:vsd} and  on some more general details relating to the computations.

  \subsection{Regression setup}\label{subsection:regsetup}
For animals in the data set we use   the  first 600 ms of the recording corresponding to 977 images. Thus for each trial  of VSD data we have a total of $250112$ observations arranged in a  $25\times 25\times 977$ grid. We model a delay of $L\coloneqq 50$ images corresponding to $\tau\approx31$ ms which gives us   $ M\coloneqq 926$ modeled time points.

\begin{figure}[H]
\begin{center}
\includegraphics[scale = 0.32]{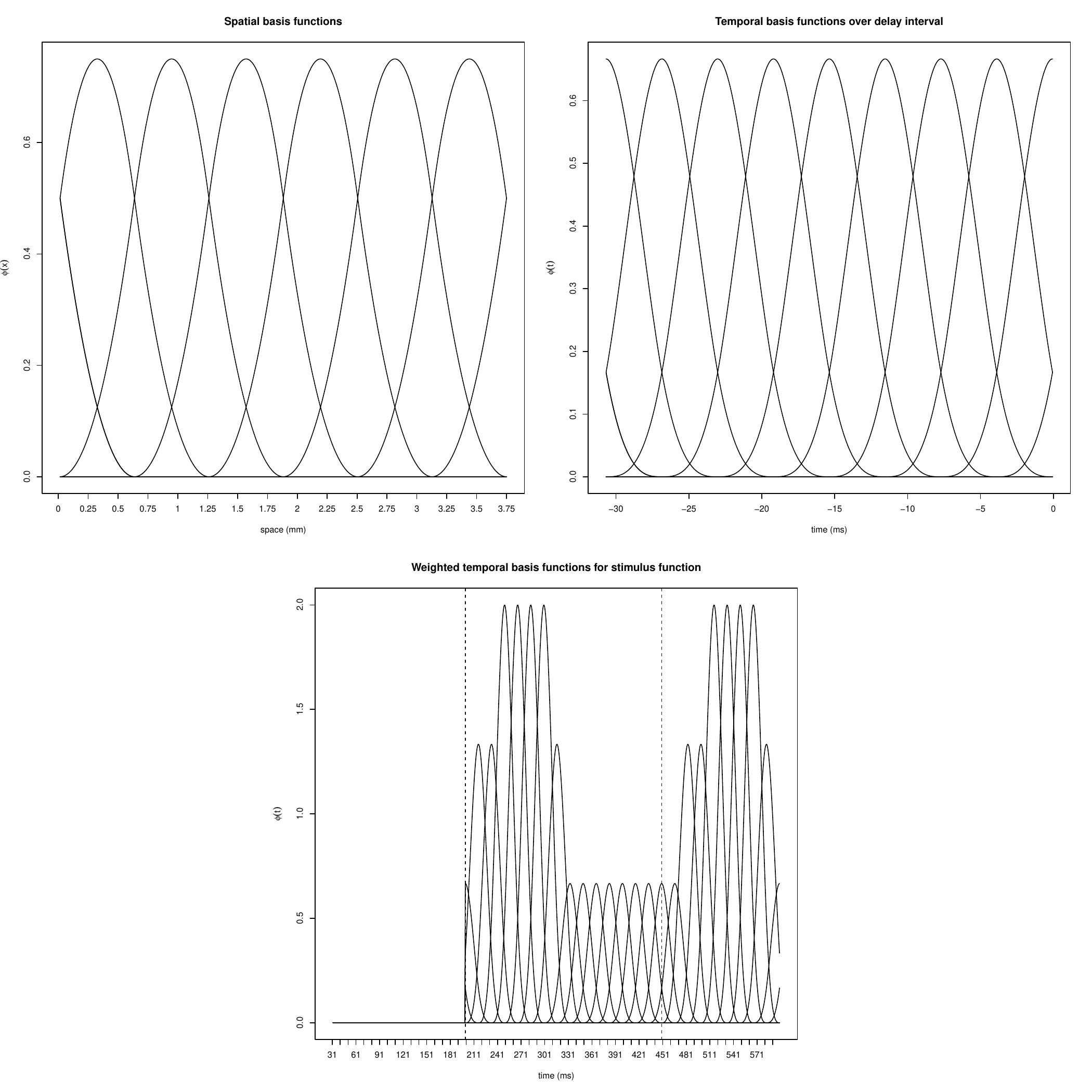}
\caption{Spatial  basis splines 
and temporal basis splines
.}
\label{fig:nine}
\end{center}
\end{figure}

As explained above we can use  tensor basis functions to represent the component functions.
For the analysis of the brain image data  we  will use   B-splines as basis functions in each dimension as these have compact support.
Specifically in the spatial dimensions we will use quadratic B-splines while in the temporal dimension we will use cubic  B-splines, see Figure \ref{fig:nine}. Note that for the temporal factor of the stimulus function we choose basis function covering  the stimulus interval and post stimulus interval.
   The number of basis functions in each  spatial dimension is $p_x=p_y\coloneqq 8$  and  in the temporal dimension we have   $p_\ell \coloneqq 11$ basis functions to capture the delay and  $p_t\coloneqq 27$ temporal basis functions to model the stimulus, see Figure \ref{fig:nine}.  With this setup we have  a total of  of $p \coloneqq p_s+p_w+p_h=p_xp_yp_t+p_xp_yp_xp_yp_\ell +p_xp_y= 46848$ model parameters that need to be estimated.   
 
For the lasso regression we give all parameters the same  weight 1, except for the   basis function representing the temporal component of the stimulus.  In particular,  basis functions located right after stimulus start and  stop are penalized less than  the rest of the basis function.  Figure \ref{fig:nine} (right) shows the basis function weighted with the inverse of the parameter weights.  With this weighing scheme the stimulus component will be focused on picking up the direct stimulus effect while it  will be less likely to pick up propagation effects.

We  fit the model for a sequence of  penalty parameters  $\lambda_{max}\coloneqq \lambda_{1}>\ldots>\lambda_{K}=:\lambda_{min}$ with  $K\coloneqq 10$. Here given data, $\lambda_{max}$ is   the smallest value yielding a zero solution to \eqref{eq414}. The results presented in the text are from model number $6$. See the supplementary material, Section 1, for more on model selection.

 \subsection{Algorithmic details}
We will here sketch how to  solve the subproblem \eqref{eq414} using a generalized block relaxation algorithm and the \textsf{R} software package \verb+glamlasso+, see \cite{lund2018}.  The entire  method, which relies on Algorithm \ref{alg:br}, is available from CRAN via  the \textsf{R}  package \verb+dynamo+, see \cite{lund2018c}.

 Algorithm \ref{alg:br} works by fixing  different blocks of the parameter in an alternating fashion and then optimize over the non fixed blocks. Let $O_\lambda\coloneqq \ell_X+\lambda J_1,\lambda>0$ denote the objective function in \eqref{eq414} considered as a function of the parameter arrays (or blocks) $\{\ar\alpha,\ar\beta,\ar{\gamma}\}$.  Let $\hat Y_{-b} $ denote the partial residuals array obtained by setting  parameter block $b\in\{\ar\alpha,\ar\beta,\ar \gamma \}$, to zero when computing the linear predictor in \eqref{eq4.15},  for given estimates of the remaining parameter blocks, and then using this to compute the model residuals.
 
Each subproblem \eqref{c1}-\eqref{c2} in Algorithm \ref{alg:br} can be solved with a function from the  \verb+glamlasso+ package. Especially \eqref{c1} is solved with \verb+glamlassoRR+,  \eqref{c2} is solved with  \verb+glamlasso+  and \eqref{c3} is solved with  \verb+glamlassoS+. These functions  are based on the coordinate descent proximal gradient algorithm, see \cite{lund2017}. We note that  the function \verb+glamlassoRR+,  performing the  reduced rank regression  in this tensor setup, also utilizes a block relaxation scheme to obtain the desired  factorization of the parameter array $\ar{ \alpha}$.   Furthermore we note that when iterating the steps 1-3 in the approximate MCRE algorithm above we have to reweigh data with the estimated covariance matrix. This in turn leads to  weighted estimation problems  in each subproblem  \eqref{c1}-\eqref{c2}.

\begin{algorithm}[h]
                \caption{Block relaxation algorithm}
                \label{alg:br}
                \begin{algorithmic}
                                \REQUIRE $ \{\ar \alpha^{(0)},\ar \beta^{(0)}, \ar\gamma ^{(0)}\}$, $ \Phi^{xy\ell},\Phi^t,\Phi^x,\Phi^y, \hat \Sigma $
                                \WHILE{$a<N\in \NN$}
                                \STATE $a=a+1$
                                \FOR{$b=1$ to $3$}
                                \IF{$b = 1$}
                                \STATE compute $\hat Y_{-\ar \alpha}$ using $\ar{\hat \beta}^{(a)}, \ar{\hat \gamma} ^{(a)}$
                                and  solve the reduced  rank problem
                                \begin{alignat}{4}\label{c1}
                                \ar{ \hat \alpha}^{(a+1)}=\arg\min_{\ar \alpha} O_\lambda(\ar\alpha,\ar{ \hat \beta}^{(a)}, \ar{\hat\gamma} ^{(a)})
                                \quad \text{s.t.} \quad \ar \alpha_{i,j,k}=\zeta_k\eta_{i,j}
                                \end{alignat}  
                                using  data $\hat Y_{-\ar \alpha}$, $\Phi^x,\Phi^y,\Phi^t, \hat \Sigma$.
                                \ELSIF{$b = 2$}
                                \STATE compute $\hat Y_{-\ar \beta}$ using $\ar{ \hat \alpha}^{(a+1)}, \ar{\hat \gamma} ^{(a)},$ and     solve  
                                \begin{alignat}{4}\label{c2}
                                \ar{ \hat \beta}^{(a+1)}=\arg\min_{\ar \beta} O_\lambda(\ar{ \hat \alpha}^{(a +1)}, \ar \beta,\ar{\hat \gamma} ^{(a)})
                                \end{alignat}
                                using       data $\hat Y_{-\ar \beta}$, $\Phi^x,\Phi^y,\Phi^{xy\ell}, \hat \Sigma$.
                                \ELSIF{$b = 3$}
                                \STATE compute $\hat Y_{-\ar \gamma }$ using $\ar{ \hat \alpha}^{(a+1)}, \ar{ \hat \beta}^{(a +1)}$ and     solve 
                                \begin{alignat}{4}\label{c3}
                                \ar{ \hat \gamma} ^{(a+1)}=\arg\min_{\ar \gamma } O_\lambda(\ar{ \hat \alpha}^{(a +1)},\ar{ \hat \beta}^{(a +1)}, \ar \gamma )
                                \end{alignat}
                                using   data $\hat Y_{-\ar \gamma }$, $\Phi^x,\Phi^y, \hat \Sigma$.
                                \ENDIF
                                \ENDFOR
                                \IF{convergence criterion is satisfied}
                                \STATE break
                                \ENDIF
                                \ENDWHILE
                \end{algorithmic}
\end{algorithm}

\subsection{Computing the convolution tensor} 
We note that for the filter component, the  tensor component $\Phi^{xy\ell}$, as shown in the proof of Proposition \ref{prop:three}, corresponds to a   convolution of the random field.  This component has to be computed upfront which in principle could be very time consuming. However considering \eqref{eqa5}  this computation can  be carried out using array arithmetic. Especially  we can write the convolution tensor as
\begin{alignat*}{4}
\Phi^{xy\ell} =
\left(
\begin{array}{ccc}
\ve(\Phi^{xy\ell}_{1})
\\
\vdots
\\
\ve(\Phi^{xy\ell}_{M})
\end{array}
\right),
\end{alignat*}
where   $\Phi^{xy\ell}_{k}$ for each $k$ is a  $p_x\times p_y\times p_\ell $ array which according to \eqref{eqa5} can  be computed using the array arithmetic as
\begin{alignat*}{4}
\Phi^{xy\ell}_{k} \coloneqq  
\rho[(\phi^\ell)^\top, \rho\{(\phi^y)^\top, \rho((\phi^x)^\top,   (\ar{\tilde {\tilde V}}_\ell)_{\ell=k-L}^{k-1}) \}].
\end{alignat*}

\bibliography{bibliotek}
\includepdf[pages = -]{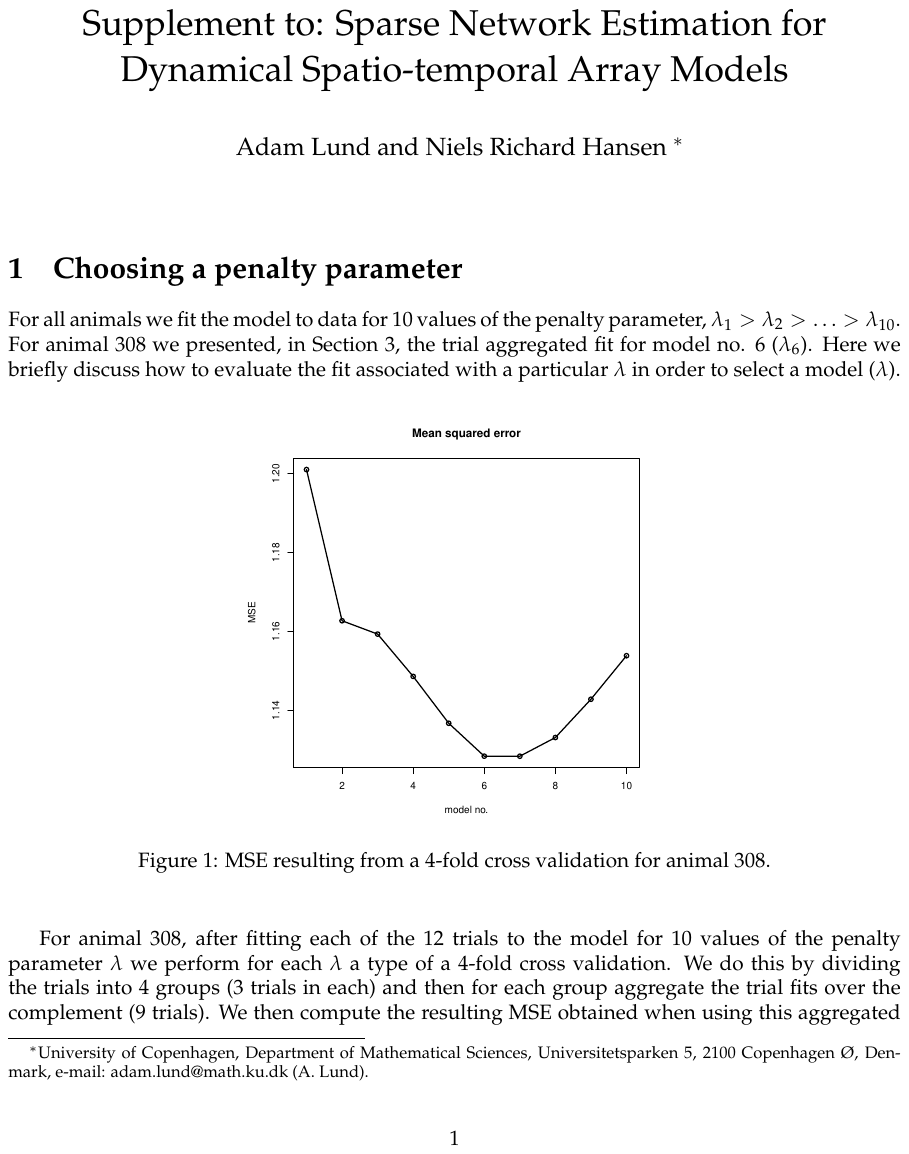}
\end{document}